\documentclass[prb,showpacs,amsmath,amssymb,superscriptaddress,twocolumn,
floatfix]{revtex4}

\usepackage[english]{babel}
\usepackage{graphicx}
\usepackage{times}
\usepackage{units}

\begin{document}

\title{Van Hove singularities in the paramagnetic phase of the Hubbard
model: a DMFT study}

\author{Rok \v{Z}itko}
\affiliation{Jo\v{z}ef Stefan Institute, Jamova 39, SI-1000 Ljubljana, Slovenia}

\author{Janez \surname{Bon\v ca}}
\affiliation{Faculty of Mathematics and Physics, University of
Ljubljana, Jadranska 19, SI-1000 Ljubljana, Slovenia}
\affiliation{Jo\v{z}ef Stefan Institute, Jamova 39, SI-1000 Ljubljana, Slovenia}

\author{Thomas Pruschke}
\affiliation{Institute for Theoretical Physics, University of G\"ottingen,
Friedrich-Hund-Platz 1, D-37077 G\"ottingen, Germany}
\affiliation{Racah Institute of Physics, The Hebrew University
of Jerusalem, Jerusalem, Israel}

\date{\today}


\begin{abstract}
Using the dynamical mean-field theory (DMFT) we study the paramagnetic phase
of the Hubbard model with the density of states (DOS) corresponding to the
three-dimensional cubic lattice and the two-dimensional square lattice, as
well as a DOS with inverse square root singularity. We show that the
electron correlations rapidly smooth out the square-root van Hove
singularities (kinks) in the spectral function for the 3D lattice and that
the Mott metal-insulator transition (MIT) as well as the
magnetic-field-induced MIT differ only little from the well-known results
for the Bethe lattice. The consequences of the logarithmic singularity in
the DOS for the 2D lattice are more dramatic. At half filling, the
divergence pinned at the Fermi level is not washed out, only its integrated
weight decreases as the interaction is increased. While the Mott transition
is still of the usual kind, the magnetic-field-induced MIT falls into a
different universality class as there is no field-induced localization of
quasiparticles. In the case of a power-law singularity in the DOS at the
Fermi level, the power-law singularity persists in the presence of
interaction, albeit with a different exponent, and the effective impurity
model in the DMFT turns out to be a pseudo-gap Anderson impurity model with
a hybridization function which vanishes at the Fermi level. The system is
then a generalized Fermi liquid. At finite doping, regular Fermi liquid
behavior is recovered.
\end{abstract}

\maketitle

\newcommand{\korr}[1]{\langle\langle #1 \rangle\rangle}
\renewcommand{\Im}{\mathrm{Im}}
\renewcommand{\Re}{\mathrm{Re}}
\newcommand{\vc}[1]{\mathbf{#1}}

\section{Introduction}

For many materials, it is permissible to consider each electron as moving
essentially independently in a static periodic effective potential which
takes into account the interactions between the electrons. This point of
view received theoretical support through the density-functional theory
(DFT) \cite{hohenberg1964} and the Kohn-Sham Ansatz \cite{kohn1965}, which
consists of replacing the full many-body problem with an auxiliary
independent-particle problem. DFT calculations are remarkably accurate for
wide-band systems, such as simple metals, many semiconductors and
insulators, but they are less appropriate for strongly correlated electron
systems, such as some transition metals, lanthanides and their compounds
\cite{stewart1984, imada1998}.

For non-interacting systems, the density of states (DOS) counts the number
of the single-particle levels that may be occupied per unit energy. It can
be defined as 
\begin{equation}
\label{rho0}
\rho_0(\omega) = \frac{1}{N} \sum_\vc{k} \delta(\omega-\epsilon_\vc{k}),
\end{equation}
where $\vc{k}$ indexes the $N$ single-particle levels with energies
$\epsilon_\vc{k}$. For an infinite system, the sum goes into an integral
\begin{equation}
\rho_0(\omega) = \int \frac{d^d\vc{k}}{(2\pi)^d} \delta(\omega-\epsilon_\vc{k}) =
\int \frac{d^{d-1}\vc{k}}{(2\pi)^d} \frac{1}{|\nabla
\epsilon_k|_{\epsilon_k=\omega}}.
\end{equation}
Any smooth periodic function must have critical points where the gradient
vanishes. In a periodic structure such as a crystal, $\epsilon_\vc{k}$ is
periodic in the reciprocal space, therefore $\rho_0(\omega)$ will necessarily
have singularities arising from minima, maxima and saddle points of
$\epsilon_\vc{k}$. For topological reasons, a certain minimum number of
these van Hove singularities must be present in any band structure
\cite{vanhove1953}. In two-dimensional systems, for example, the saddle
point in the dispersion gives rise to a logarithmic divergence in the DOS.
The van Hove singularities are thought to be particularly important for the
physics of low-dimensional systems, for example in oxide superconductors
\cite{newns1992, gofron1994, lu1996, markiewicz1997, irkhin2002}.

For strongly-interacting systems, the concept of the single-particle levels
is not very useful and one should resort to the techniques from the
many-particle theory. In particular, the equivalent of the DOS is the
local spectral function
\begin{equation}
\label{rho1}
\rho(\omega) = -\frac{1}{\pi} \Im\left[ G_{\mathrm{loc}}(\omega+i\delta) \right],
\end{equation}
where $G_\mathrm{loc}(\omega)$ is the local ($\vc{k}$-averaged) Green's
function 
\begin{equation}
G_{\mathrm{loc}}(\omega) = \frac{1}{N} \sum_\vc{k} G_{\vc{k}}(\omega),
\end{equation}
with $G_{\vc{k}}(z) = \korr{ c_{\vc{k}} ; c_{\vc{k}}^\dag }_z$ the
momentum-resolved Green's function (electron propagator).  In the absence of
interactions, $\Im\left[ G_{\vc{k}}(\omega+i\delta) \right] = -\pi
\delta(\omega-\epsilon_{\vc{k}})$ and Eq.~\eqref{rho1} reverts to
Eq.~\eqref{rho0}.  The interactions modify the propagation of electrons,
thus the spectral function $\rho(\omega)$ differs significantly from the
non-interacting DOS $\rho_0(\omega)$ and, in particular, any sharp features
such as van Hove singularities will smooth out due to broadening (finite
life-time) effects.

The presence of a diverging DOS at the Fermi level enhances instabilities
towards various ordered states, in particular antiferromagnetism and
ferromagnetism \cite{hlubina1997}, as well as superconductivity
\cite{hirsch1986}. In addition, the underlying paramagnetic metal phase
itself is expected to have unusual properties. In this work, we address this
regime using the dynamical mean-field theory (DMFT) \cite{georges1996},
using the numerical renormalization group (NRG) as the impurity solver
\cite{wilson1975, bulla2008, resolution}. We will study the spectral
function of the Hubbard model \cite{hubbard1963, kanamori1963,
gutzwiller1963} in the paramagnetic phase on the three-dimensional simple
cubic lattice (which has square root singularities at the edges of the band
and two further square root singularities inside the band) and on the
two-dimensional square lattice (where a logarithmic singularity is located
in the center of the band). We focus on the effects (in particular on
metal-insulator transitions \cite{imada1998}) where local physics plays the
key role, thus the use of the DMFT as an approximate method to study
finite-dimensional systems is justified. For completeness, we also consider
the case of a DOS with an integrable power-law singularity, to wit an
inverse square root DOS. While such divergences arise in 1D problems, we
consider here this case using the DMFT purely out of academic interest.

\section{Method}

In the limit of infinite dimensions or infinite lattice connectivity, the
self-energy in the Hubbard model becomes purely local \cite{metzner1989,
mullerhartmann1989, georges1996}. The bulk problem of correlated electrons
then maps exactly onto a quantum impurity model with a self-consistently
defined non-interacting bath of conduction electrons, in this case an
Anderson impurity model \cite{georges1992, rozenberg1992, jarrell1992,
zhang1993}. To solve the effective impurity model, various non-perturbative
techniques can be used, for example the numerical renormalization group
(NRG) \cite{wilson1975, krishna1980a, sakai1989, costi1994, bulla1998,
bulla1999, pruschke2000, bulla2008} which is particularly suitable to study
the low-temperature limit. The NRG calculations in this work were performed
for the discretization parameter $\Lambda=2$, with the $z$-averaging over 8
values of the twist parameter \cite{frota1986, campo2005} using a modified
discretization scheme from Ref.~\onlinecite{resolution}. Spectral functions
were computed using the density-matrix approach \cite{hofstetter2000} and
the self-energy trick \cite{bulla1998}. The truncation cutoff was set at
$E_\mathrm{cutoff}=10\omega_N$ and the broadening was performed with
parameter $\alpha=0.1$. 

The input to a NRG calculation step in the DMFT cycle is an effective
hybridization function $\Gamma_\sigma(\omega)$ which contains full
information about the coupling between the interacting impurity site and the
effective non-interacting medium. The output, as required for the DMFT
calculation, is the self-energy function $\Sigma_\sigma(\omega)$. The local
lattice Green's function is then
\begin{align}
G_{\mathrm{loc},\sigma}(\omega) &= \frac{1}{N} \sum_k G_{k,\sigma}(\omega) \\
&= \frac{1}{N} \sum_k
\frac{1}{\left[\omega+\mu_\sigma-\Sigma_\sigma(\omega)\right]-\epsilon_{k}} \\
&= \int \frac{\rho_0(\epsilon) d\epsilon}{\left[
\omega+\mu_\sigma-\Sigma_\sigma(\omega) \right]-\epsilon},\\
&= G_0 \left[ \omega+\mu_\sigma-\Sigma_\sigma(\omega) \right],
\end{align}
where $\rho_0(z)$ is the density of states in the non-interacting model,
while $G_0(z)$ is the associated free-electron propagator. The local lattice
spectral function is then
\begin{equation}
\rho_\sigma(\omega) = -\frac{1}{\pi} \Im \left[
G_{\mathrm{loc},\sigma}(\omega+i\delta) \right].
\end{equation}
The self-consistency condition \cite{georges1996} relates the local lattice
Green's function $G_{\mathrm{loc},\sigma}$ and the hybridization function
$\Gamma_\sigma$ as
\begin{equation}
\label{Gamma}
\Gamma_\sigma(\omega) = -\Im \left[
\omega-
\left(
G_{\mathrm{loc},\sigma}^{-1} +
\Sigma_\sigma(\omega)\right) \right].
\end{equation}
The DMFT iteration proceeds until two consecutive solutions for the local
spectral function differ by no more than some chosen value. To accelerate
the convergence to the self-consistency and to stabilize the solutions, one
can make use of the Broyden mixing \cite{broyden}.

\section{3D cubic lattice DOS} 

An analytical expression is known for the Green's function for the
three-dimensional simple cubic lattice \cite{joyce1972}:
\begin{equation}
\label{dos3d} 
\begin{split}
G_0(z) &= \frac{1}{z} \frac{\sqrt{1-\frac{3}{4}x_1}}{1-x_1}
\left[ \frac{2}{\pi} K(k^2_+) \right] \left[ \frac{2}{\pi} K(k^2_-) \right],
\\ k_{\pm}^2 &= \frac{1}{2} \pm \frac{1}{4} x_2 \sqrt{4-x_2}
-\frac{1}{4}(2-x_2)\sqrt{1-x_2}, \\ x_1 &= \frac{1}{2} +
\frac{1}{6}z^2-\frac{1}{2} \sqrt{1-z^2} \sqrt{1-\frac{1}{9}z^2}, \\
x_2 &= \frac{x_1}{x_1-1},
\end{split}
\end{equation}
where $K(m)$ is the complete elliptic integral of the first kind. When using
these expressions, one should be careful to choose the correct convention
for the argument of the elliptic integrals, since several different are in
common use. In the dimensionless form of Eq.~\eqref{dos3d}, the band is
centered at zero and the bandwidth is equal to 6. The behavior of the DOS at
the band edges is typical for three-dimensional systems: the DOS goes to
zero as $\sqrt{\Delta \omega}$, where $\Delta \omega$ is the distance from
the band edge. The DOS is thus continuous, however its derivative diverges.
This feature is shared by the DOS of the Bethe lattice with infinite
connectivity which is, for this very reason, a common model DOS for three
dimensional systems. Inside the band there are two further square root
singularities where the DOS is continuous, but the derivatives are
discontinuous and diverging on one side.

\begin{figure}[htbp]
\includegraphics[width=8cm,clip]{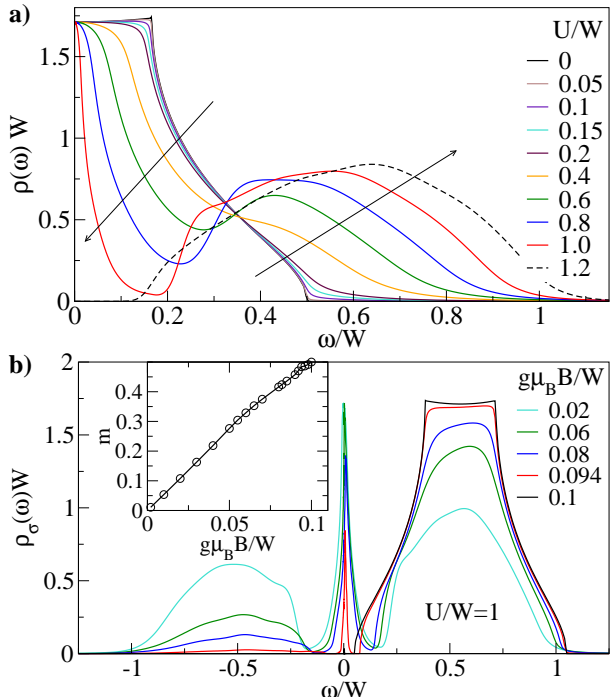}
\caption{(Color online) a) Spectral functions for the Hubbard model with the
3D cubic lattice DOS in the paramagnetic phase at half-filling, $T=0$. Due
to the particle-hole symmetry only positive frequencies are shown. The
arrows indicate the direction of increasing interaction $U$. 
The dashed curve corresponds to a result for the insulating phase.
b) The evolution of the spectral function as a function of the magnetic
field. The inset shows the field-dependence of the magnetization.
}
\label{fig1}
\end{figure}

All four van Hove singularities smooth out as soon as the interaction is
turned on, see Fig.~\ref{fig1}a. At moderate $U/W \sim 0.6$ the spectral
function already strongly resembles that obtained for a structureless DOS
such as the one for the Bethe lattice with infinite coordination
\cite{bulla1999}. Nevertheless, the shape of the non-interacting DOS does
have an effect on the quantitative details. The Mott-Hubbard metal-insulator
(MIT) transition occurs at a value of $U_\mathrm{c}/W = 1.165$, which is to
be compared with the results for the Bethe lattice, $U_\mathrm{c}/W=1.47$
\cite{bulla1999}. If the result is rescaled in terms of the effective
bandwidth defined through the second moment of the DOS \cite{bulla1999}
\begin{equation}
W_\mathrm{eff} = 4 \sqrt{ \int_{-W/2}^{W/2} d\epsilon\, \epsilon^2
\rho(\epsilon) } \approx 0.816 W,
\end{equation}
we obtain $U_\mathrm{c}/W_\mathrm{eff} = 1.43$, which compares well with the
Bethe-lattice result. The feature that $W_\mathrm{eff}$ is a rather robust
characteristic preferable to the bandwidth has been noted before
\cite{bulla1999}.

The Hubbard model undergoes a magnetic-field-induced MIT in external
magnetic field, the main mechanism on the Bethe lattice being a
field-induced quasiparticle mass enhancement (field-induced localization)
\cite{laloux1994, georges1996, bauer2009fm}. 
The behavior for a 3D cubic lattice DOS is very similar: the quasiparticle
peak narrows down and moves slightly away from the Fermi level, while the
Hubbard bands become increasingly spin-polarized and they take the form of
the non-interacting DOS, however they do not move much, see
Fig.~\ref{fig1}b. The MIT occurs when the quasiparticle peak vanishes. This
happens for a field on the scale of the width of the quasiparticle peak in
the absence of the field. For $U/W=1$, for example, we obtain $g\mu_B
B_\mathrm{c}=0.099W$, which is to be compared with the quasiparticle residue
$Z=0.077$.

\section{2D square lattice DOS} 

\begin{figure}[htbp]
\includegraphics[width=7cm,clip]{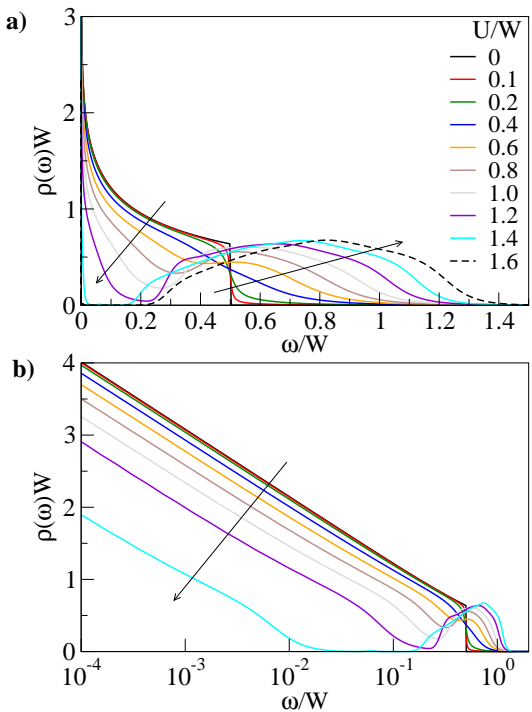}
\caption{(Color online) Spectral functions for the Hubbard model with the 2D
square lattice DOS in the paramagnetic phase at half-filling, plotted on a
a) linear and b) logarithmic energy scale. The arrows show the direction of
increasing $U$. The result plotted with dashed curve is already in the
paramagnetic insulator phase.}
\label{new1}
\end{figure}

The free-electron propagator on the two-dimensional square lattice is
\cite{economou}
\begin{equation}
G_0(z)=\frac{2}{\pi z} K\left( \frac{1}{z^2} \right),
\end{equation}
with a $z \to 0$ expansion
\begin{equation}
G_0(z)=\frac{\pi+i(2\ln z-4\ln2)}{2\pi} + \mathcal{O}(z^2\ln z),
\end{equation}
which gives the logarithmic singularity in the density of states at the
Fermi level for a half-filled system:
\begin{equation}
\rho_0(\omega) \sim -\frac{2\ln\omega-4\ln2}{2\pi^2}.
\end{equation}
The divergence at the Fermi level in the DOS is {\em not} eliminated as the
interaction is turned on, see Fig.~\ref{new1}. The effect of the interaction
at low energy scales is to renormalize the constant part $4\ln2/(2\pi^2)$ to
smaller values, while the logarithmic term keeps the same prefactor. We find
that the logarithmic scaling is difficult to achieve numerically to very low
energies, where spurious features were observed for energies below
$10^{-6}W$. In spite of the diverging spectral function, the Mott
metal-insulator transition appears to be of the usual type. With increasing
interaction $U$, a region of low spectral density appears before the onset
of the logarithmic peak. We find $U_\mathrm{c}/W_\mathrm{eff}=1.45$, in
agreement with the standard result (it should be noted that for the 2D
square lattice DOS we have $W_\mathrm{eff}=W$). The value of $U_\mathrm{c}$
was determined by studying the sharp resonances in $\Im \Sigma(\omega)$ in
the metal phase which evolve into the zero-frequency pole in the self-energy
for the insulator phase \cite{zhang1993}. We extracted their position $X$ as
a function of the interaction $U$, and solved for $U_\mathrm{c}$ in the
equation $X(U_\mathrm{c})=0$.

For strong interaction, the Hubbard bands have a strongly asymmetric shape
with pronounced shoulders near the inner band edges, and less pronounced
shoulders at the outer band edges. These features survive into the
insulating phase, which is to be contrasted with the behavior in the case of
Bethe and 3D cubic lattice, where the structure at the inner band edge
disappears as the transition point is approached \cite{karski2005,
resolution}. The nature of the shoulders observed here is thus different:
they reflect the discontinuities at the band edges of the non-interacting DOS.

\begin{figure}[htbp]
\includegraphics[width=8cm,clip]{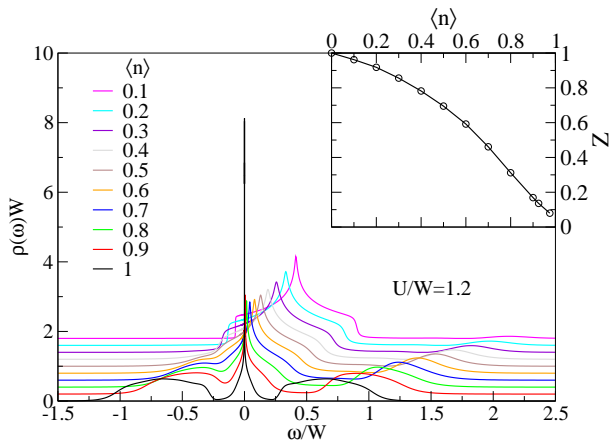}
\caption{(Color online) Spectral functions for the Hubbard model with the 2D
square lattice DOS for different doping levels. The curves are offset
vertically for clarity. The inset shows the quasiparticle residue
as a function of the doping.}
\label{new2}
\end{figure}

When the non-interacting system is doped, the logarithmic singularity moves
away from the Fermi level. Albeit the energy resolution of NRG is finite at
energies away from the Fermi level, it is nevertheless sufficient to study
the interaction-induced broadening (the energy resolution of NRG at finite
energies has been recently studied in Ref.~\onlinecite{resolution}). A plot
of the spectral functions as a function of doping is shown in
Fig.~\ref{new2}. At finite doping, the imaginary part of the self-energy is
finite at the energy of the impurity level, thus the singularity transforms
into an asymmetric Lorentzian-like peak. The quasiparticle residue 
(wavefunction renormalization)
\begin{equation}
Z=\left(1-\partial \Sigma(\omega)/\partial \omega\right)^{-1}
\end{equation}
goes to one with increased doping. It should be noted that at half-filling,
$\Re\Sigma(\omega)$ has a diverging slope, thus $Z \to 0$. The system is
thus not a regular Fermi liquid, but rather a singular Fermi liquid
\cite{varma2002}.

\begin{figure}[htbp]
\includegraphics[width=8cm,clip]{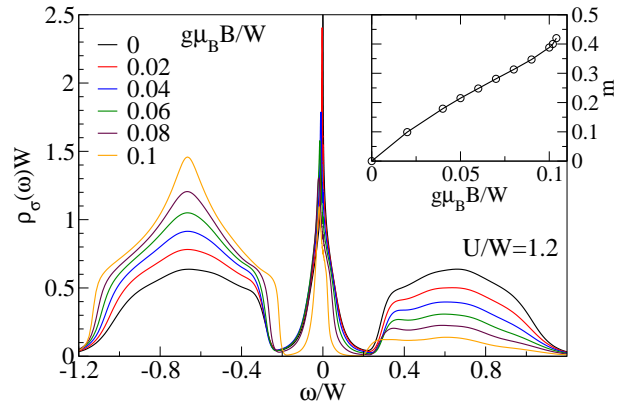}
\caption{(Color online) Spectral functions of the Hubbard model with the 2D
square lattice DOS in the magnetic field; half-filling case. 
The inset shows the field-dependence of the magnetization.}
\label{new3}
\end{figure}

The system exhibits unusual properties in the magnetic field. At
half-filling, the behavior of the quasiparticle peak with increasing field
is different from that in the systems with finite DOS at the Fermi level; 
it grows {\em broader} rather than narrower, while its integrated weight
decreases, see Fig.~\ref{new3}. As the transition point is approached from
below, we find that the DMFT calculations no longer converge even when the
advanced Broyden mixing is used. This signals the complete {\em absence} of
solutions to the DMFT equations, rather than merely their instability or
metastability. This might signal the presence of an intermediate phase which
must be qualitatively different from both the homogeneous
partially-spin-polarized paramagnetic phase and the fully polarized
insulating phase.

\begin{figure}[htbp]
\includegraphics[width=7cm,clip]{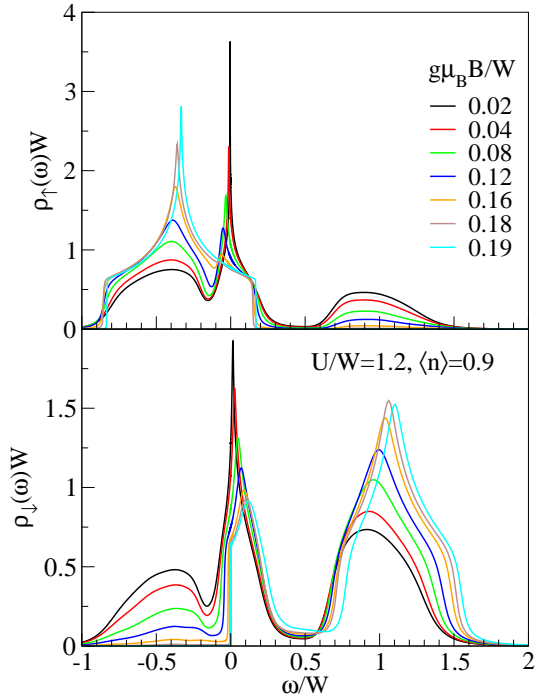}
\caption{(Color online) Spectral functions of the Hubbard model
with the 2D square lattice DOS in the magnetic field; finite-doping
case.}
\label{new4}
\end{figure}

At finite filling, the behavior in the magnetic field becomes more in line
with that of systems with non-diverging DOS, see Fig.~\ref{new4}. With
increasing field, the quasiparticle peak splits. The majority-spin resonance
is reduced in amplitude and eventually disappears as the lower Hubbard band
becomes increasingly non-interacting-like. At the same time, the
minority-spin resonance becomes wider and transforms into a feature with a
sharp edge near the Fermi level. The spin-minority upper Hubbard band is
significantly renormalized even for relatively strong magnetic fields. Such
behavior is analogous to that found for the Bethe lattice
\cite{bauer2007fm}. 

\section{DOS with power-law singularity}

\begin{figure}[htbp]
\includegraphics[width=7cm,clip]{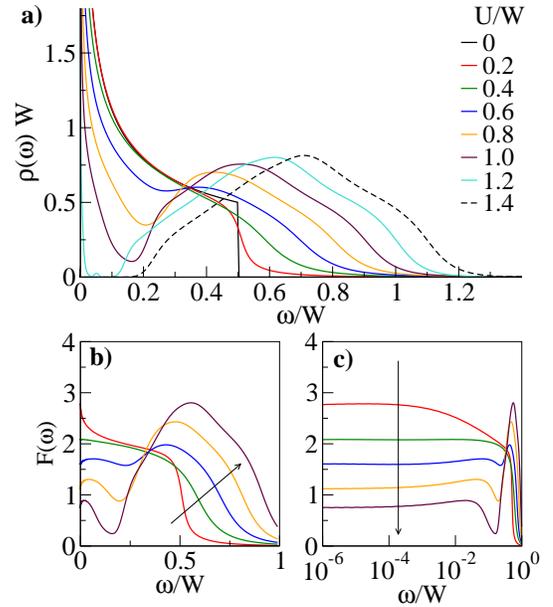}
\caption{(Color online) a) Spectral functions for the Hubbard model with the
power-law DOS $|\omega|^{-1/2}$ in the paramagnetic phase at half-filling.
b,c) Rescaled spectral functions on linear and logarithmic energy scale. The
arrows show the direction of increasing $U$.}
\label{fig2}
\end{figure}

Motivated by the unusual features induced by the presence of a logarithmic
(i.e. rather mild) singularity at the Fermi level in the DOS of the 2D
square lattice, we now study the case of stronger power-law singularities on
the example of an inverse square root DOS:
\begin{equation}
\rho_0(\omega) = \frac{1}{2\sqrt{2} W} \left|\omega/W\right|^{-1/2}.
\end{equation}
The DMFT results indicate that in the presence of the interaction, the
power-law singularity remains, however its exponent changes: the spectral
functions feature a $\omega^{-\alpha}$ singularity with the exponent around
$\alpha \approx 0.40$. The energy below which this power-law scaling of the
spectral spectral function holds depends exponentially on the value of $U$
for small $U$, while for $U \gtrsim 0.1 W$, it starts to hold essentially on
the scale of bare parameters (i.e. $U$ itself). In Fig.~\ref{fig2}b,c we
plot the modified spectral functions \cite{glossop2000}
\begin{equation}
F(\omega)=\pi \sec^2\left( \frac{\pi}{2} \alpha \right) |\omega|^\alpha
\rho(\omega),
\end{equation}
which reveal the structure of the power-law singularity. 
The decrease of $F(0)$ with increasing $U$ corresponds to the progressive
narrowing of the quasiparticle peak as the MIT is approached. The transition
is found to occur for $U_\mathrm{c}/W=1.32$ or
$U_\mathrm{c}/W_\mathrm{eff}=1.48$, again in good agreement with the standard
result.  As for other DOS functions, the Mott MIT proceeds by the same route
\cite{moeller1995, bulla1999}: with increasing $U$ a region of reduced
spectral density appears, while the weight of the quasiparticle peak
decreases until the peak disappears at the critical $U_\mathrm{c}$. 

\begin{figure}[htbp]
\includegraphics[width=6cm,clip]{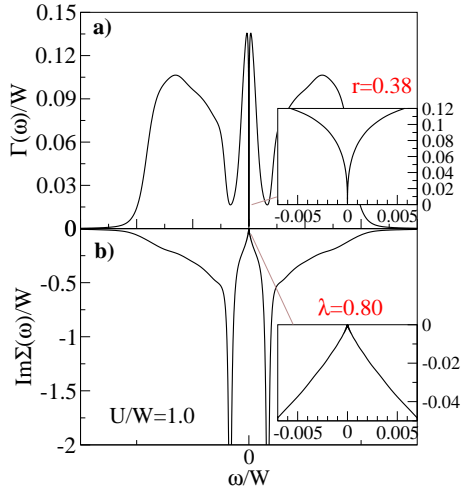}
\caption{(Color online) The hybridization function and the imaginary part of
the self-energy for the Hubbard model with the power-law DOS 
$|\omega|^{-1/2}$ in the paramagnetic phase at half-filling, $T=0$. The
insets are close-ups on the low-energy region.}
\label{fig3}
\end{figure}

Interestingly, at half filling the effective quantum impurity model is a
pseudo-gap Anderson model \cite{withoff1990, chen1995, ingersent1996,
bulla1997, gonzalez1998, glossop2000, bulla2000, logan2000,
vojta2002eurpjb}: the hybridization function $\Gamma(\omega)$, shown in
Fig.~\ref{fig3}a, has a power-law pseudogap $|\omega|^r$ with the exponent
$r=0.38$. This is a direct consequence of diverging spectral function and
the imaginary part of the self-energy $\Im \Sigma(\omega)$ going to 0 at
$\omega=0$. Since $\Im\Sigma(\omega)$ has a cusp-like $\omega^\lambda$
singularity with the exponent $\lambda \approx 0.80$, see Fig.~\ref{fig3}b,
the self-energy goes to zero faster than the hybridization function ($r <
\lambda$), thus the system can be classified as a generalized Fermi liquid
in the sense of Refs.~\onlinecite{gonzalez1998} and
\onlinecite{glossop2000}. Within numerical accuracy, the spectral function
exponent $\alpha$ is equal to the hybridization exponent $r$, as expected
\cite{logan2000, bulla2000}. 

The low-energy expansion of the self-energy can be expressed as
\cite{glossop2000}
\begin{equation}
\Sigma(\omega) \propto -|\omega|^\lambda \left[i+\tan\left(\lambda (\pi/2)
\right)\mathrm{sgn}\omega\right]
\end{equation}
Since the inverse spectral function goes to zero faster than $\omega$ and
the self-energy $\Sigma(\omega)$, the self-consistency condition
Eq.~\eqref{Gamma} reduces to
\begin{equation}
\Gamma(\omega)=\Im\left[ G_\mathrm{loc}^{-1}(\omega) \right].
\end{equation}
Furthermore, taking into account the non-interacting DOS, the local Green's
function can be expressed as
\begin{equation}
G_\mathrm{loc}(\omega) = G_0\left[ \omega-\Sigma(\omega) \right],
\end{equation}
which reduces to $G(\omega)=G_0\left[ -\Sigma(\omega) \right]$, since
$\Sigma(\omega)$ goes to zero slower than $\omega$. If $G_0\left[z\right]$
is the free-electron propagator in a system with power-law DOS
$|\omega|^{-R}$, the low-$\omega$ expansion of the hybridization function is
found to be 
\begin{equation}
\Gamma(\omega) \propto \left|\omega\right|^{\lambda R}.
\end{equation}
In our case with $R=1/2$, we have $\lambda=0.8$ and the numerically obtained
exponent $r=0.38$ agrees well with the expected value of $r=\lambda R =
0.4$. This implies that $r<\lambda$ for all $R \in [0:1]$ for which the
singularity is integrable, thus a strong-coupling fixed point (rather than
local-moment fixed point) is expected in general and the system will
always be a generalized Fermi liquid. Furthermore, the perturbation theory
in $U$ for the pseudo-gap Anderson impurity model gives a relation
$\lambda=2-3r$ \cite{glossop2000}, thus we obtain a result for the exponent
of the self-energy
\begin{equation}
\lambda=\frac{2}{1+3R},
\end{equation}
which is expected to hold in general. For $R=1/2$, this yields
$\lambda=4/5$, which is corroborated by our numerical results.

\begin{figure}[htbp]
\includegraphics[width=8cm,clip]{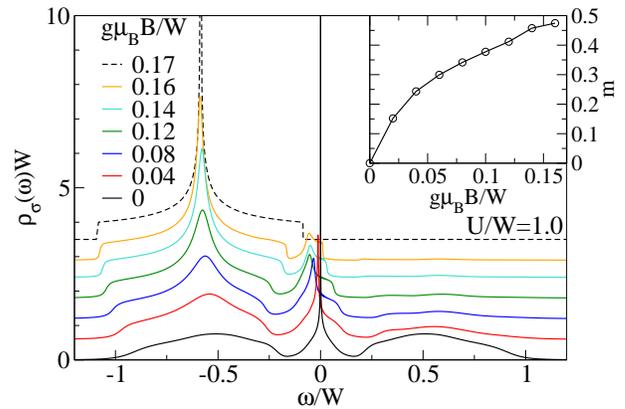}
\caption{(Color online) The spectral function of the Hubbard model with the
power-law DOS $|\omega|^{-1/2}$ in the magnetic field. The curves are offset
vertically for clarity. The inset shows the field-dependence of the
magnetization.}
\label{fig4}
\end{figure}

The behavior in the magnetic field is very different from that found in
previously known cases, see Fig.~\ref{fig4}. With increasing field, the
diverging quasiparticle peak transforms into a finite peak slightly away
from the Fermi level and its width {\it grows} with field. At the same time,
the majority-spin lower Hubbard band becomes increasingly
non-interacting-like with emerging van Hove singularities in its center and
sharp band edges. The MIT is not driven by quasi-particle mass enhancement
and vanishing width of the quasiparticle band. Instead, it occurs when the
edge of the quasiparticle band crosses the Fermi level.

\begin{figure}[htbp]
\includegraphics[width=8cm,clip]{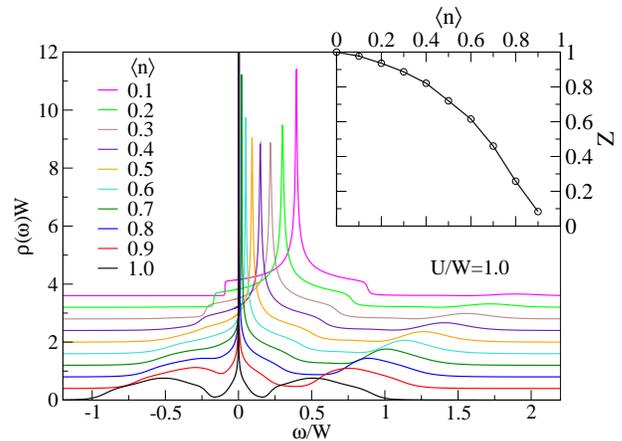}
\caption{(Color online) Spectral functions for the Hubbard model with the 
power-law DOS $|\omega|^{-1/2}$ for different doping levels. The curves are
offset vertically for clarity. The inset shows the quasiparticle residue as
a function of the doping.}
\label{fig2n}
\end{figure}

Upon doping, the quasiparticle peak transforms smoothly into the van Hove
singularity in the center of the empty band, see Fig.~\ref{fig2n}. At finite
doping, the effective hybridization function $\Gamma(\omega)$ no longer
attains zero value at $\omega=0$, however it still exhibits a sharp
pseudo-gap feature with a minimum which is shifted away from the Fermi level
and where $\Gamma(\omega)>0$. For small doping the minimum still appears
cusp-like, however it becomes increasingly parabolic-like for larger doping.
The evolution of the quasiparticle residue as a function of the doping is
shown in the inset to Fig.~\ref{fig2n}. At half-filling, the system is a
generalized Fermi liquid, thus $Z=0$. At finite doping, the system is a
genuine Fermi liquid with a self-energy which behaves as $\Im\Sigma \sim
\omega^2$ near the Fermi level. 

\section{Conclusion}

We have shown that the systems with a non-interacting density of states
which diverges at the Fermi level behave as singular Fermi liquids, since
the singularities are not washed out by the interactions. Their
distinguishing characteristic is the different behavior in the magnetic
field, in particular the absence of field-induced quasiparticle
localization. This has implications for the possibility of fully polarizing
such systems with external magnetic fields of the order of the width of the
quasiparticle band.

\begin{acknowledgments}
This work has been supported by DFG collaborative research center, SFB 602,
Schonbrunn Fellowship of the Hebrew University, and Gesellschaft f\"ur
wissenschaftliche Datenverarbeitung (GWDG). TP acknowledges the hospitality
of the Racah Institute of Physics.
\end{acknowledgments}

\bibliography{paper}

\end{document}